\documentclass{ws-ijmpcs}

\graphicspath{ {figs/} }
\usepackage{epsfig}

\allowdisplaybreaks

\begin{document}

\markboth{M. Steinhauser}
{Towards analytic $(g-2)_\mu$ at 4 loops}

%
\catchline{}{}{}{}{}
%

\title{Towards analytic $(g-2)_\mu$ at four loops}


\author{Matthias Steinhauser
}

\address{Institut f{\"u}r Theoretische Teilchenphysik, Karlsruhe
  Institute of Technology (KIT),\\ 76128 Karlsruhe, Germany\\
matthias.steinhauser@kit.edu}

%

\maketitle


\begin{abstract}
  In this contribution we present recent four-loop
  results for the muon anomalous magnetic moment 
  based on analytic methods. In particular,
  fermionic corrections involving two or more closed 
  electron loops or at least one tau lepton loop are discussed.

\end{abstract}

\ccode{PACS numbers: 12.20.-m 12.38.Bx 14.60.Cd 14.60.Ef}

\section{Introduction}

The anomalous magnetic moment of the muon constitutes both
experimentally and theoretically a clean quantity which makes it an ideal
object for precision studies. The current
experimental value for $a_\mu=(g-2)_\mu/2$
(see Refs.~\refcite{Bennett:2006fi,Roberts:2010cj})
\begin{eqnarray}
  a_{\mu}^{\rm exp} &=& 116\,592\,089(63) \times 10^{-11}
  \,,
  \label{eq::amuexp}
\end{eqnarray}
matches the accuracy of the theoretical
prediction which is given by~\cite{Hagiwara:2011af} (see also
Refs.~\refcite{Jegerlehner:2011ti,Davier:2010nc})
\begin{eqnarray}
  a_{\mu}^{\rm th} &=& 116\,591\,828(49) \times 10^{-11}
  \,.
  \label{eq::amuth}
\end{eqnarray}
However, the comparison of Eqs.~(\ref{eq::amuexp}) and~(\ref{eq::amuth}) 
shows that there is a discrepancy of about 3$\sigma$ which 
persists for more than a decade.

The numerically most important contribution to $a_{\mu}^{\rm th}$ comes from
QED followed by hadronic, light-by-light and electroweak corrections which are
described in detail in the
reviews~\refcite{Melnikov:2006sr,Jegerlehner:2009ry}
and~\refcite{Miller:2012opa}.  In this contribution recent four-loop QED
corrections~\cite{Lee:2013sx,KLMS13} are discussed which are based on
analytic calculations with the purpose to provide an independent check of
the purely numerical approach of Ref.~\refcite{Aoyama:2012wk}. 
One of the motivations for such a cross-check is the fact that the
four-loop QED contribution is of the same order of magnitude as the difference
between Eqs.~(\ref{eq::amuexp}) and~(\ref{eq::amuth}).  Beyond one-loop order
large QED corrections are obtained from Feynman diagrams containing closed
electron loops.  Since the electron mass cannot be set to zero such diagrams
lead to sizeable logarithms $\ln(M_\mu/M_e) \approx 5.3$ which
occur up to third power at four loops. Among the various classes of Feynman
diagrams those where the external photon couples to a closed electron loop,
the so-called light-by-light-type diagrams, give the most important
contributions. At three loops, where analytic results are known, these
diagrams come with an additional factor $\pi^2$.  In our approach the
light-by-light diagrams are technically quite demanding and have not yet been
considered. As a preparatory work we looked at other classes
which are discussed in the following two sections: the
contribution involving closed tau lepton loops (see Section~\ref{sec::tau})
and the one with two or three closed electron loops (see
Section~\ref{sec::electron}).  The corresponding results have been obtained in
Refs.~\refcite{KLMS13} and~\refcite{Lee:2013sx}.

\section{\label{sec::tau}Closed tau loops}

\begin{figure}[t]
  \begin{center}
      \includegraphics[scale=0.65]{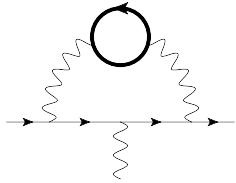}
  \end{center}
  \vspace*{-1em}
  \caption[]{\label{fig::diag_2l} Two-loop Feynman diagrams contributing to
    $(g-2)_\mu$.  Thin and thick solid lines represent muon and tau leptons,
    respectively, and wavy lines denote photons.}
\end{figure}

\begin{figure}[t]
  \begin{center}
      \leavevmode
      \epsfxsize=0.9\textwidth
      \epsffile[80 450 500 750]{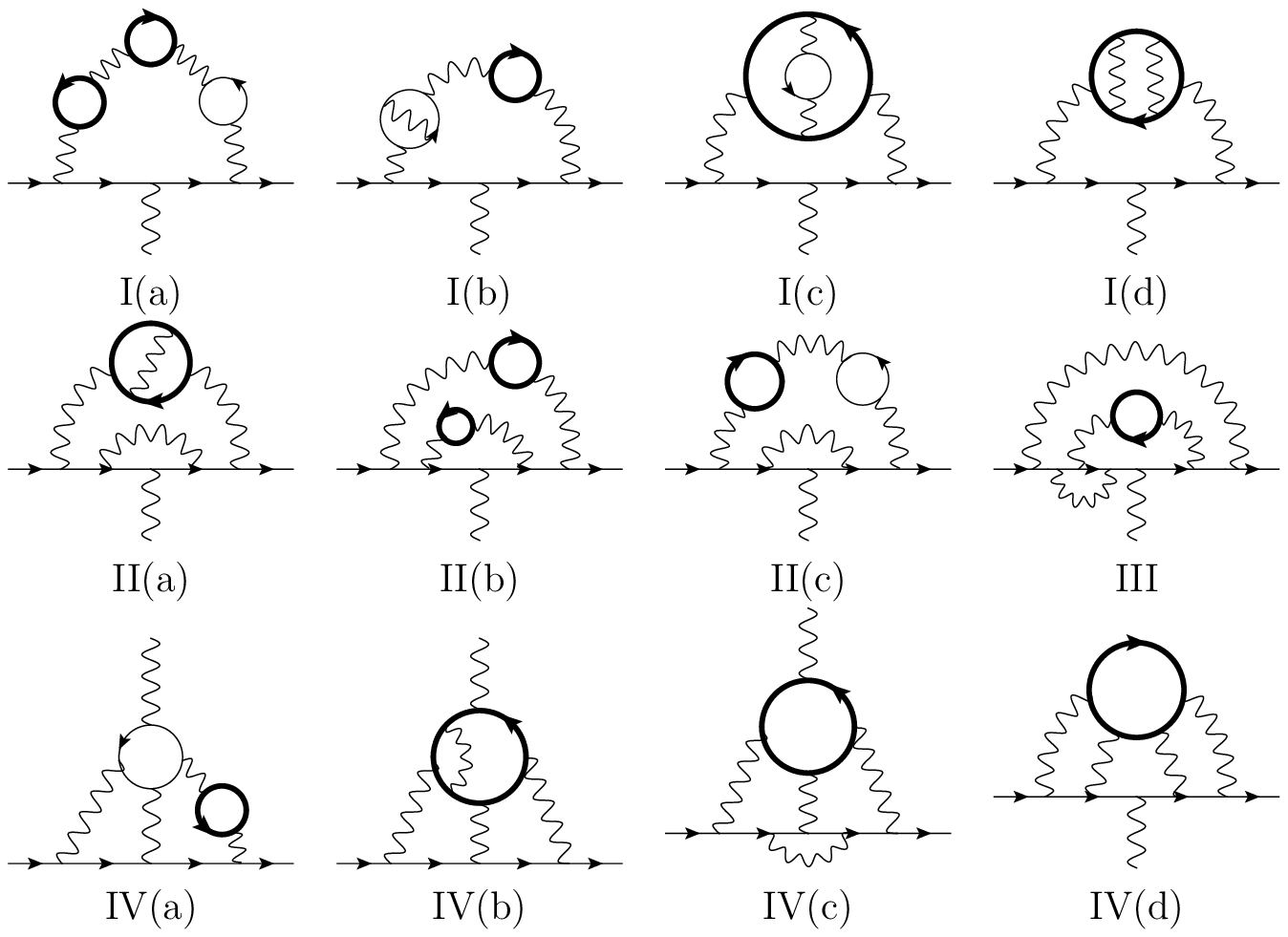}
  \end{center}
  \vspace*{-1em}
  \caption[]{\label{fig::diags_4l} Sample Feynman diagrams contributing to
    $(g-2)_\mu$ at four-loop order.  The symbols 
    label the individual diagram classes and are taken over from
    Refs.~\refcite{Aoyama:2012wj,Aoyama:2012wk}.}
\end{figure}

Starting from two loops there are Feynman diagrams contributing to $a_\mu$
which contain a closed tau lepton loop. Actually there is only one such
diagram at two-loop order
(see Fig.~\ref{fig::diag_2l}) since the contributions where the
external photon couples to the closed fermion loop are zero due to Furry's
theorem.  At three loops one has to deal with 60 and at four loops with 1169
Feynman diagrams.  The four-loop diagrams can be subdivided into twelve
classes~\cite{Aoyama:2012wk} which are shown in Fig.~\ref{fig::diags_4l}.

\begin{figure}[t]
  \begin{center}
    \includegraphics[scale=0.55]{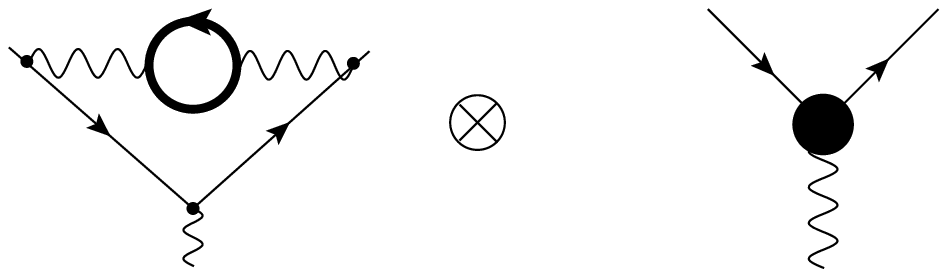}
    \hfill \includegraphics[scale=0.55]{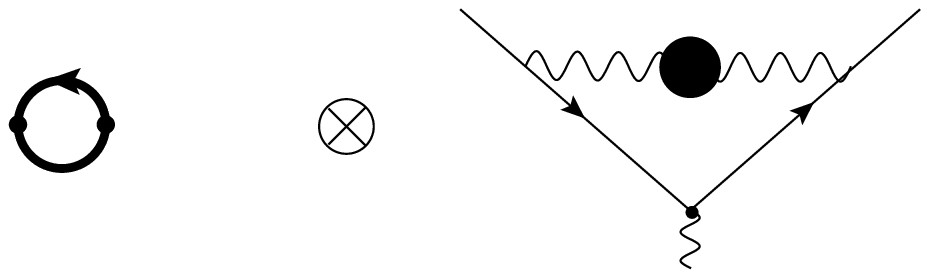}
  \end{center}
  \caption[]{\label{fig::ae} Graphical representation of 
    the hard sub-graphs and co-subgraphs as obtained after applying
    the rules for asymptotic expansion to the two-loop diagram
    in Fig.~\ref{fig::diag_2l}.}
\end{figure}

The two-loop diagram can be computed exactly~\cite{Elend:1966} 
in terms of a function which depends on $M_\mu/M_\tau$ (see also
Refs.~\refcite{Laporta:1992pa,Czarnecki:1998rc,Passera:2006gc}).
We nevertheless want to use this simple example in order to
demonstrate the method which we apply at four loops where an exact
calculation is out of reach with the currently available technology. The basic
idea is to obtain an expansion of $a_\mu$ in the limit $M_\tau \gg M_\mu$ by
Taylor expanding the integrand in certain kinematical regions. The latter is
visualized in Fig.~\ref{fig::ae} where the two contributions are shown which
arise after applying the rules of asymptotic expansion.\cite{Smirnov:2013} The
notation is as follows: left of the symbol $\otimes$ one finds the so-called
hard subgraphs which by definition contain all tau lepton propagators and which are
one-particle irreducible with respect to the light lines. The subgraphs are
expanded in all small quantities, in our case the external momenta and the
muon mass, and afterwards the integrations over the hard loop momenta are
performed.  On the right of $\otimes$ one finds the co-subgraphs. They are
constructed from the original diagram by removing all lines which are part
of the subgraph.  The blob indicates the position 
where the result of the subgraph has to
be inserted before integration over the loop momenta of the co-subgraph.  
At two-loop order the hard subgraphs lead to
either one- or two-loop vacuum integrals with one mass scale, $M_\tau$,
whereas for the co-subgraphs
tree-level contributions or one-loop on-shell integrals with $q^2=M_\mu^2$
have to be considered.  
Analogously, at four
loops one has to deal with vacuum integrals up to four loops and on-shell
integrals up to three loops. Both classes of integrals are very well studied
up to this loop-order (see Ref.~\refcite{KLMS13} for references and
more details).

In Ref.~\refcite{KLMS13} three expansion terms in $M_\mu^2/M_\tau^2\approx
0.0035$ have been computed for all twelve classes of diagrams shown in
Fig.~\ref{fig::diags_4l}.  After adding all contributions one obtains
\begin{eqnarray}
  A^{(8)}_{2,\mu}(M_\mu/M_\tau)&\approx&
  0.0421670 + 0.0003257 + 0.0000015 \approx 0.0424941(2)(53)
  \,,
  \label{eq::A8mu}
\end{eqnarray}
where $(\alpha/\pi)^4 A^{(8)}_{2,\mu}(M_\mu/M_\tau)$ represents the four-loop
contribution to $a_\mu$ induced by virtual tau lepton loops.  The first and
second uncertainty in Eq.~(\ref{eq::A8mu}) indicates the truncation error and
the error in the input quantity $M_\mu/M_\tau$, respectively.  Due to the
smallness of the expansion parameter we observe a rapid convergence. Actually,
as can be seen after the first approximation sign in Eq.~(\ref{eq::A8mu}) each
subsequent term is about a factor 100 smaller than the previous one. Thus it
is save to assign 10\% of the last computed term as uncertainty of the
truncation after $(M_\mu^2/M_\tau^2)^3$.  Comparing the result
of Eq.~(\ref{eq::A8mu}) with the one from Ref.~\refcite{Aoyama:2012wk}
based on numerical integration, $0.04234(12)$, shows good agreement. However,
the analytic result is significantly more precise.

\section{\label{sec::electron}Closed electron loops}

\begin{figure}[t]
  \begin{center}
      \includegraphics[scale=0.75]{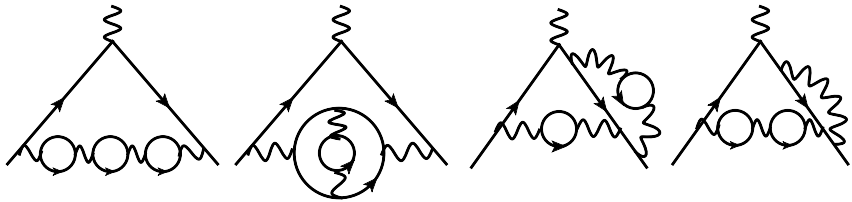}
  \end{center}
  \vspace*{-1em}
  \caption[]{\label{fig::amu_e} Four-loop Feynman diagrams contributing to
    $(g-2)_\mu$. In at least two of the closed fermion loops electrons are
    present.}
\end{figure}

In Ref.~\refcite{Lee:2013sx} a first step towards a systematic study of four-loop
on-shell integrals has been undertaken. More precisely, all classes of Feynman
integrals have been studied which are needed to compute QED or QCD corrections
to a massive fermion propagator with on-shell external momenta and two or
three closed massless loops. Thus, contributions to $a_\mu$ from diagrams as
shown in Fig.~\ref{fig::amu_e} can be evaluated.

The four-loop integrals considered in Ref.~\refcite{Lee:2013sx} only contain
either massive of massless lines. For this reason, as a start, the electrons
have to be chosen as massless which leads to finite results as long as the
fine structure coupling is renormalized in a mass independent renormalization
scheme. Thus, initially, we renormalize the muon mass on-shell but $\alpha$ in
the $\overline{\rm MS}$ scheme.  The (finite) result contains both constant
terms and logarithms $\ln(\mu^2/M_\mu^2)$ where $\mu$ is the renormalization
scale of the fine structure constant.  Afterwards we transform $\alpha$ to
the on-shell scheme which introduces $\ln(\mu^2/M_e^2)$ terms. In this way the
$\mu$ dependence cancels and $\ln(M_e^2/M_\mu^2)$ terms remain.

By construction the described approach can only be applied to 
those Feynman diagrams where the closed electron loops
are related to the renormalization of $\alpha$. This excludes the 
light-by-light-type Feynman diagrams where the external photon couples to an
electron.

As an example we want to discuss the four-loop contribution to $a_\mu$ which
contains two closed electron loops but no additional closed muon loop. In
Ref.~\refcite{Lee:2013sx} this contribution has been denoted by
$a_\mu^{(42)a}$ and the analytic expression is given by\footnote{For similar
  corrections of this type see also
  Refs.~\refcite{Laporta:1993ds,Aguilar:2008qj}.} 
\newcommand{\Mu}{M_\mu} \newcommand{\Me}{M_e}
\newcommand{\lnue}{L_{\mu e}}
\begin{eqnarray}
  a_\mu^{(42)a} &=&
    \lnue^2
    \left[\pi ^2
      \left(\frac{5}{36}-\frac{a_1}{6}\right)+\frac{\zeta_3}{4}
      -\frac{13}{24}\right]  
   + \lnue \left[-\frac{a_1^4}{9}+\pi ^2
    \left(-\frac{2 a_1^2}{9}+\frac{5
    a_1}{3}-\frac{79}{54}\right)
    \right.\nonumber\\&&\left.\mbox{}
    -\frac{8 a_4}{3}-3 \zeta_3+\frac{11 \pi ^4}{216}
    +\frac{23}{6}\right]
   -\frac{2 a_1^5}{45}+\frac{5 a_1^4}{9}
    +\pi ^2 \left(-\frac{4
    a_1^3}{27}+\frac{10 a_1^2}{9}
    \right.\nonumber\\&&\left.\mbox{}
    -\frac{235
    a_1}{54}-\frac{\zeta_3}{8}+\frac{595}{162}\right)
    +\pi ^4 \left(-\frac{31
    a_1}{540}-\frac{403}{3240}\right)+\frac{40 a_4}{3}+\frac{16
    a_5}{3}-\frac{37 \zeta_5}{6}
    \nonumber\\&&\mbox{}
    +\frac{11167 \zeta_3}{1152}-\frac{6833}{864}
    \,,
    \label{eq::amu_e}
\end{eqnarray}
with $a_1=\ln2$, $a_n=\mbox{Li}_n(1/2)$ ($n\ge1$), $\zeta_n$ is Riemann's zeta
function and $\lnue=\ln(\Mu^2/\Me^2)$. It is interesting to note that
quantities up to trancendentality level five appear in Eq.~(\ref{eq::amu_e}).  
The numerical evaluation leads to
$a_\mu^{(42)a} = -3.62427$ which should be compared to
$-3.64204(112)$.\cite{Kinoshita:2004wi,Aoyama:2012wk}  The difference is of
order $10^{-2}$ (i.e. $0.5\%$) and can be explained by missing
$M_e/M_\mu$ terms in the analytic expression.

\section*{Acknowledgments}
I would like to thank Alexander Kurz, Roman Lee, Tao Liu, Peter Marquard,
Alexander Smirnov and Vladimir Smirnov for a fruitful collaboration on the
topics discussed in this contribution.  This work was supported by the DFG
through the SFB/TR~9 ``Computational Particle Physics''.

\vspace*{-1em}


\end{document}